\documentclass[12pt]{iopart}
\usepackage{amssymb}
\usepackage{iopams}
\usepackage{setstack}
\usepackage{amstext}
\usepackage{graphicx}
\usepackage[section]{placeins}
\usepackage{relsize}
 
\begin{document}
 
\title{Current loops and fluctuations in the zero-range process on a
diamond lattice}
\author{R Villavicencio-Sanchez, R J Harris and H Touchette}
\address{School of Mathematical Sciences, Queen Mary University of
London, Mile End Road,
London, E1 4NS, UK.}
\eads{\mailto{sanchez@maths.qmul.ac.uk},
\mailto{rosemary.harris@qmul.ac.uk}, \mailto{ht@maths.qmul.ac.uk}}
 
\begin{abstract}
We study the zero-range process on a simple diamond lattice with open boundary conditions and determine the conditions for the existence of loops in the mean current. We also perform a large deviation analysis for fluctuations of partial and total currents and check the validity of the Gallavotti-Cohen fluctuation relation for these quantities. In this context, we show that the fluctuation relation is not satisfied for partial currents between sites even if it is satisfied for the total current flowing between the boundaries. Finally, we extend our methods to study a chain of coupled diamonds and demonstrate co-existence of mean current regimes.
\end{abstract}
 
\section{Introduction}
 
In the last decades much effort has been put into understanding and
modelling non-equilibrium systems,
which find applications in fields ranging from biology to finance
\cite{BlyMcK,BouPot}. Among the various
models which have been proposed to study such real-life processes,
stochastic interacting particle systems
(interacting Markov systems) have enjoyed particular success
\cite{BerSolGab}. In this class, the zero-range
process (ZRP) is a well studied lattice gas model offering many
applications and the possibility of obtaining
analytical results. Introduced in 1970 \cite{Spi}, one of the reasons
the ZRP gained interest was because
it can show a phase transition from a fluid to a condensed state
\cite{EvaHan}. The ZRP has been extensively
studied with both periodic and open boundary conditions in one
dimension \cite{Eva00,LevMukSch05}, and some
variants involving junction topologies have also been introduced, see
e.g. \cite{AngSchZia, MarBer}.
Furthermore, currents in a closely related model have recently been
considered on more general networks
\cite{CheCheGol}, which may give some insight into expected effects
for manmade networks such as traffic on
roads or the internet.\\
 

For extended ZRPs defined on two-dimensional or three-dimensional
lattices, particle currents can flow in principle
in loops within the bulk of the system. To illustrate this
possibility, we study a simple variant of the ZRP defined
on a diamond-shaped lattice with open boundaries. For this model, we
are able to obtain analytical expressions for
the mean current flowing between specific sites of the lattice which
show that the model has two regimes of mean
particle currents: a unidirectional regime where particles flow in the
same direction through different lattice
branches and a loop current regime where the mean current flows around
the diamond. The analysis is also extended to a chain of coupled diamonds where, 
for a weak asymmetry of the bulk hopping rates, we find a transition 
between unidirectional and loop current regimes persisting in the 
large-system-size limit.\\
 
In addition to mean currents, we also study fluctuations by
calculating the probability distribution of
particle currents and its associated large deviation rate function,
which plays a role similar to thermodynamic potentials
in equilibrium systems \cite{Tou09,Der07}. Recently much experimental and
theoretical attention has been devoted to the study of certain
fluctuation symmetries \cite{HarSch07,EspHarMuk}, which may also be
observed experimentally, e.g. \cite{JouGarCil, GarCil}. Our simple
model allows us to study explicitly the
joint probability of observing a given current on different lattice
branches. We obtain analytically
the large deviation functions for currents across different bonds and
hence gain understanding of the role of current loops.
These results for current loops, which are difficult to obtain for 
general models, are expected to be important in
testing fluctuation symmetries for higher dimensional systems,
such as those reported recently by Hurtado et al. \cite{HurPer}.\\
 
The remainder of the paper is structured as follows. In section
\ref{s:model}, we define the ZRP on the diamond lattice and
calculate its stationary state. In section \ref{s:currentloops}, we
discuss the appearance of unidirectional and loop mean current
regimes in the system. In
section \ref{s:currentfluct}, we calculate the joint particle current
fluctuations of the upper and lower branches of the
lattice and test the well known Gallavotti-Cohen fluctuation relation
for partial and total currents. In section \ref{s:diamondchain} we use 
our approach to analyse a chain of diamonds. 
Finally, in section \ref{s:conclusion}, we discuss our results and their
possible implications for other models.
 
\section{Zero-range process on a diamond lattice}\label{s:model}
 
\subsection{Definition of model}
 
\begin{figure}[h]
	\begin{center}
		\includegraphics[scale=0.7]{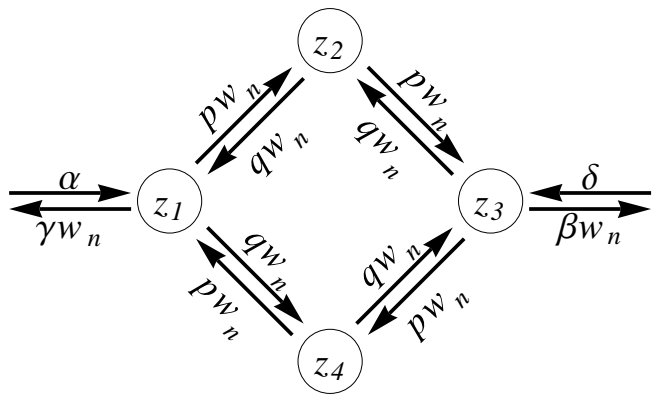}
	\end{center}
	\caption{Diamond array. The $z_i$'s represent the fugacities of sites
1--4. The quantities $\alpha$,
	$\gamma$, $\delta$, $\beta$, $p$ and $q$ are the hopping rates.}
	\label{DiaArr}
\end{figure}

The ZRP is a lattice-based many-particle model in which, as the name
suggests, particles interact only with
other particles at the same site. Additionally, particles are allowed
to accumulate to any non-negative number
on each site of the lattice. Here, we study the ZRP with open boundary
conditions on a diamond
lattice as shown in figure \ref{DiaArr}.\\
 
The dynamics of the particles on this lattice is defined in continuous
time such that the topmost particle on
each site hops to an adjacent site after an exponentially distributed
waiting time. More precisely, particles
hop clockwise around the diamond with rate $pw_n$ and anti-clockwise
with rate $qw_n$, where the interaction
between particles on each site is taken into account by the term $w_n$
which depends exclusively on the occupation
number $n$ of the departure site. Given the symmetry of the system,
without loss of generality we will assume from
now on that
\begin{equation}
	p\ge q \label{cond1}.
\end{equation}
Moreover, we allow particles to enter and leave the boundary sites
with probability rates $\alpha$ and $\beta w_n$
for site 1, $\delta$ and $\gamma w_n$ for site 3.\\
 
Based on the one-dimensional open boundary ZRP studied in
\cite{LevMukSch05} we expect the system to be driven out of
equilibrium for $\alpha/\gamma\neq\delta/\beta$
but that some choices of parameters may lead to a boundary
condensation phenomenon in which particles accumulate on one of the
sites 1 or 3. We shall return to this point
later.\\
 
In the quantum Hamiltonian formalism \cite{Sch01}, one defines a
probability vector
\begin{equation}
	|P\rangle=\sum\limits_n P(n)|n\rangle
\end{equation}
where $|n\rangle$ is a basis vector for the particle configuration
$n=(n_1,n_2,n_3,n_4)$ and $P(n)$ the probability of that
configuration.
Then the time evolution of the system is described by the master equation
\begin{equation}
 \frac{d|P\rangle}{dt}=-H|P\rangle\label{MEq}.
\end{equation}
Here the matrix $H$, or Hamiltonian, is the stochastic generator of
the system. To explicitly write the Hamiltonian of our system we
define
the creation and annihilation operators on site $i$ by
\begin{equation}
 a_i^+=\left( \begin{array}{cccc}
                0 & 0 & 0 & \quad \\
		1 & 0 & 0 & \dots \\
		0 & 1 & 0 & \quad \\
		\quad & \vdots & \quad & \ddots
              \end{array} \right)
\text{ and }a_i^-=\left( \begin{array}{cccc}
                0 & w_1 & 0 & \quad \\
		0 & 0 & w_2 & \dots \\
		0 & 0 & 0 & \quad \\
		\quad & \vdots & \quad & \ddots
              \end{array} \right)
 \end{equation}
respectively. With the additional definition of the diagonal matrix
$d_i=w_j\delta_{j,k}$, the Hamiltonian of the
model shown in figure \ref{DiaArr} is written
\begin{eqnarray}
	-H=& \alpha\left(a_1^+-1\right) + \gamma\left(a_1^--d_1\right)
	+ \delta\left(a_3^+-1\right) + \beta\left(a_3^--d_3\right)\nonumber\\
	&+ \sum\limits_{k=1}^{4}p\left(a_k^-a_{k+1}^+-d_k\right) +
q\left(a_k^+a_{k+1}^--d_{k+1}\right) \label{HamDia}
\end{eqnarray}
where the index sums are taken modulo 4.
 
\subsection{Steady state}
 
We are interested in finding the non-equilibrium stationary state
$|P^*\rangle$ of our system. By definition, this
probability does not change in time and is therefore, such that
\begin{equation}
 	H|P^*\rangle=0.
\end{equation}
It has been shown that for the ZRP on any lattice geometry the steady
state factorises as the tensor product (see, e.g., \cite{EvaMajZia})
\begin{equation}
	|P^*\rangle=|P_1^*(n_1)\rangle\otimes|P_2^*(n_2)\rangle\otimes...\otimes|P_L^*(n_L)\rangle.
\end{equation}
Here each site's probability distribution vector is given by
\begin{equation}
	|P_k^*(n_k)\rangle=\sum\limits_{n_k} P_k(n_k)|n_k\rangle,
\end{equation}
with $P_k(n_k)$ the probability of finding $n_k$ particles on the
$k$th-site and $|n_k\rangle$ the corresponding configuration vector.
Furthermore, $P_k(n_k)$ is given in terms of the site's fugacity $z_k$
and the interaction term $w_n$ by
\begin{equation}
	P_k(n_k)=Z_k^{-1}z_k^{n_k}\prod\limits_{i=1}^{n_k}w_i^{-1},
\end{equation}
where $Z$ is the grand canonical partition function
\begin{equation}
	Z_k=\sum\limits_{j=0}^{\infty}z_k^j\prod\limits_{l=1}^{j}w_l^{-1}.\label{InfSum}
\end{equation}
It now becomes obvious that the choice of the interaction $w_n$ will
be reflected in the existence of the partition function. For instance,
if we let
\begin{equation}
	\lim\limits_{n\rightarrow\infty} w_n=\kappa,
\end{equation}
with $\kappa$ a constant, then we have to make sure that $z<\kappa$ in
order for the infinite sum of equation (\ref{InfSum}) to converge and
the partition function to exist. Hopping parameters leading to
$z>\kappa$ correspond physically to a growing condensate. In the
remainder of this paper we will assume, unless stated otherwise, an
unbounded interaction rate $w_n$ for the particles, i.e.,
\begin{equation}
\lim\limits_{n\rightarrow\infty} w_n=\infty.
\end{equation}
This guarantees that the system has a non-equilibrium stationary state
without condensation.\\
 
It can be shown that the creation and annihilation operators act on
the stationary-state eigenvector of the Hamiltonian as
$a_k^+|P_k^*\rangle=z_k^{-1}d_k|P_k^*\rangle$ and
$a_k|P^*_k\rangle=z_k|P_k^*\rangle$ respectively. Here we have made
explicit that the operators corresponding to the $k$th-site act only
on the probability distribution vector of the same site. Thus,
applying the Hamiltonian (\ref{HamDia}) to the stationary state vector
$|P^*\rangle$ yields the expression
\begin{eqnarray}
 H|P^*\rangle=&-\left[ \left(\alpha - \gamma
z_1+qz_2-pz_1+pz_4-qz_1\right)z_1^{-1}d_1 \right. \nonumber\\
 &+\left(pz_1 - qz_2+qz_3-pz_2\right)z_2^{-1}d_2 \nonumber\\
 &+\left(\delta - \beta z_3+pz_2-qz_3+qz_4-pz_3\right)z_3^{-1}d_3 \nonumber\\
 &+\left(qz_1 - pz_4+pz_3-qz_4\right)z_4^{-1}d_4 \nonumber\\
 &\left. -\left(\alpha - \gamma z_1+\delta-\beta z_3\right) \right]|P^*\rangle
\end{eqnarray}
and for $|P^*\rangle$ to be the required stationary state the
coefficients of the matrices $d_i$ must vanish. Solving the resulting
system of equations leads to the fugacities
\begin{eqnarray}
z_1=\frac{(p+q) \alpha  \beta +(p^2+q^2) (\alpha +\delta )}{(p+q)
\beta  \gamma + (p^2+q^2) (\beta +\gamma)},\nonumber\\
z_2=\frac{p \alpha  \beta+q \gamma  \delta +(p^2+q^2) (\alpha +\delta
)}{(p+q) \beta  \gamma + (p^2+q^2) (\beta +\gamma)},\nonumber\\
z_3=\frac{(p+q) \gamma  \delta +(p^2+q^2) (\alpha +\delta )}{(p+q)
\beta  \gamma + (p^2+q^2) (\beta +\gamma)},\nonumber\\
z_4=\frac{q \alpha  \beta+p \gamma  \delta +(p^2+q^2) (\alpha +\delta
)}{(p+q) \beta  \gamma + (p^2+q^2) (\beta +\gamma)}\label{fugacities}.
\end{eqnarray}
Note that with these solutions, one verifies that $\alpha - \gamma
z_1+\delta-\beta z_3=0$, which is consistent with the stationary state
being the eigenvector with eigenvalue zero.\\
 
With these fugacities, we can calculate the mean time-averaged current
$\bar j_{a,b}$ which measures the average number of particles jumping
from site $a$ to site $b$ per unit time. This current is expressed in
terms of the fugacities as
\begin{equation}
	\bar j_{k,k+1}=pz_k-qz_{k+1}=-\bar j_{k+1,k}\label{mean_curr}.
\end{equation}
Here once again, $k\in\left\{1,2,3,4 \right\}$ and modulo 4 applies
over the subindex addition. Similarly, for the boundary sites the mean
current is
\begin{eqnarray}
	\bar j_{L}=\alpha-\gamma z_1,\nonumber\\
	\bar j_{R}=\beta z_3-\delta,
\end{eqnarray}
with the convention of a positive flow direction from the left to the
right boundary of the lattice.\\
 
Since we have assumed the system is in a stationary state without
condensation, the mean currents must satisfy
\begin{equation}
	\bar j_L=\bar j_{1,2}+\bar j_{1,4}=\bar j_{2,3}+\bar j_{4,3}=\bar j_R
\end{equation}
by particle conservation. From this we then find that the mean current 
at the left boundary is positive if
\begin{equation}
	\alpha-\gamma z_1>0,\nonumber\\
\end{equation}
which is equivalent to
\begin{equation}
	\frac{\left(p^2+q^2\right)\left(\alpha\beta-\gamma\delta\right)}{\left(p+q\right)\beta\gamma+\left(p^2+q^2\right)\left(\beta+\gamma\right)}>0\nonumber\\
\end{equation}
or
\begin{equation}
	\alpha\beta-\gamma\delta>0.\label{cond2}
\end{equation}
This condition implies a net current left-to-right through the system.
If conditions (\ref{cond1}) and (\ref{cond2}) are satisfied,
it follows that the largest fugacity is $z_1$. Hence, returning to the
discussion of a bounded interaction $w_n$, the consistency
condition for existence of a stationary state without condensation is
$z_1<\kappa$.
 
\section{Particle current loops}\label{s:currentloops}
 
\begin{figure}[h]
	\begin{center}
		\includegraphics[scale=0.4]{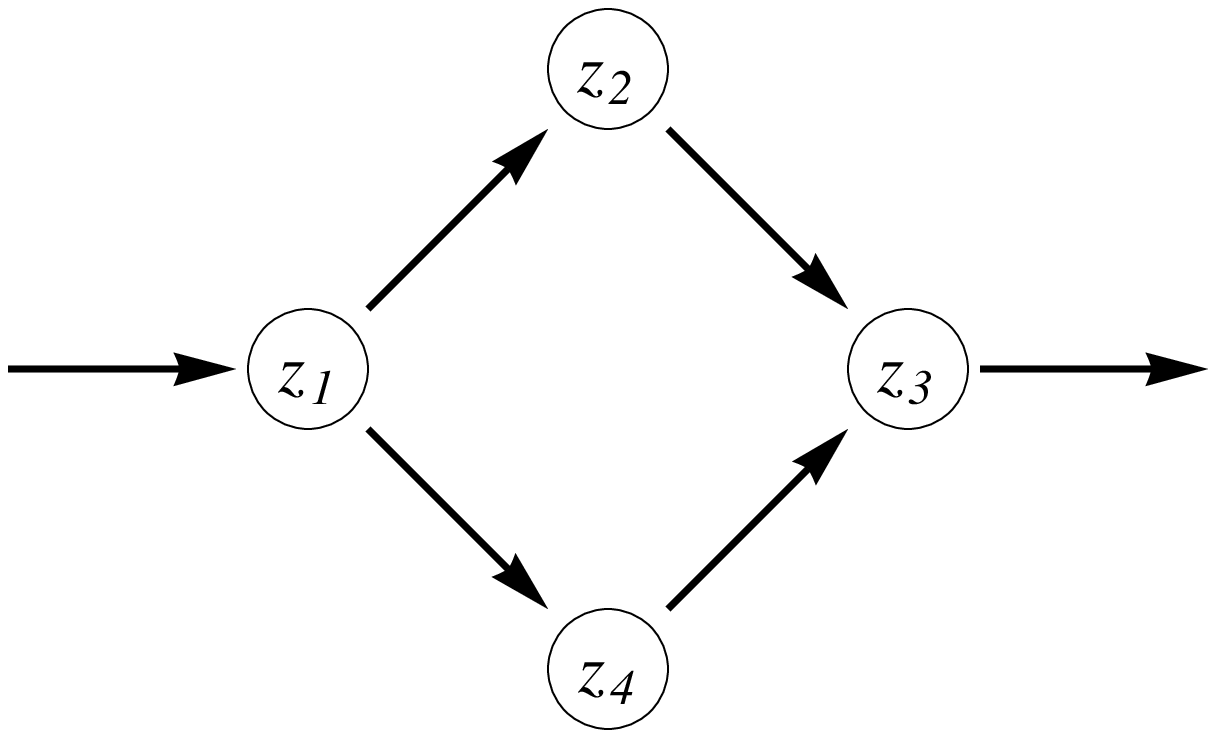}\quad\includegraphics[scale=0.4]{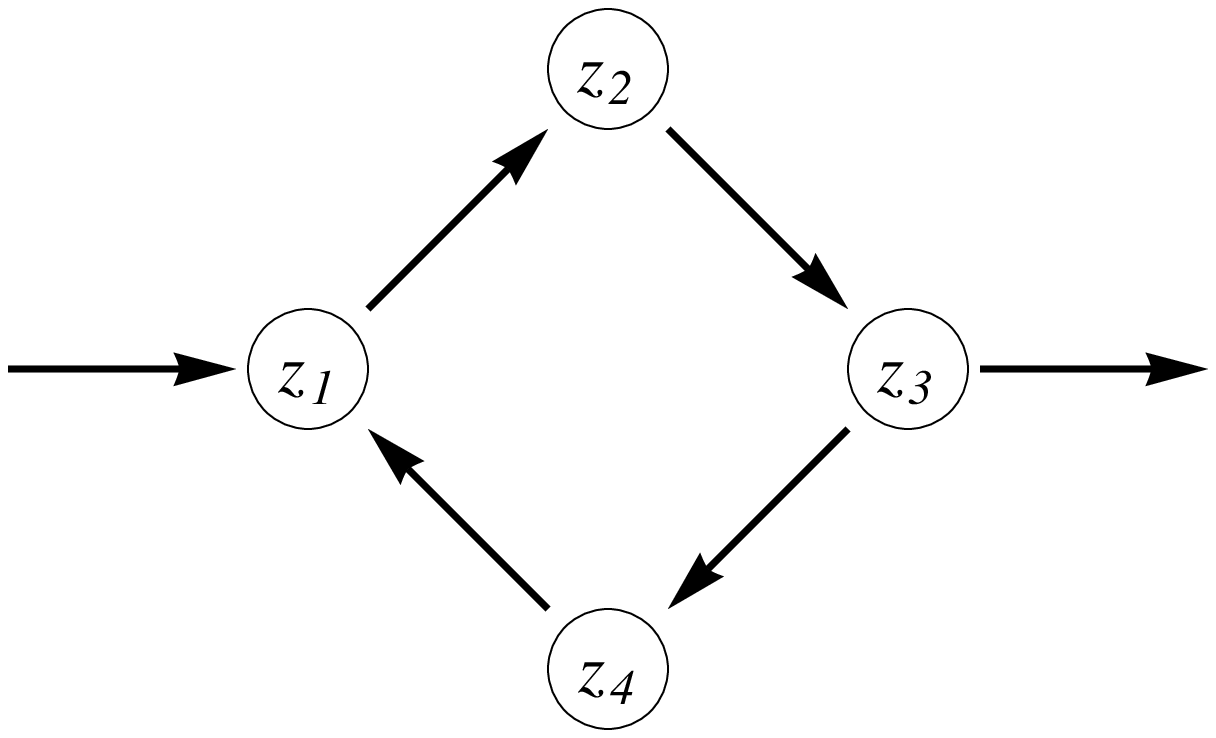}
	\end{center}
	\caption{Two regimes: Unidirectional (left) and loop mean currents (right).}
	\label{unid-loop}
\end{figure}

An interesting feature of the model we are studying is that it shows a
transition from a unidirectional mean current to a loop mean current
regime as the parameters are varied -- see figure \ref{unid-loop}. An
intuitive way to see this change, is to consider specific parameter
values. On one hand, when we have symmetric hopping rates $p=q=1$, the
mean current flows clockwise through the upper branch and
anti-clockwise through the lower branch, i.e., the unidirectional
regime. On the other hand, when we have completely asymmetric hopping
rates $p=1$ and $q=0$, the mean current flow is forced to go around
the diamond, i.e., the loop current regime. Since the mean current
through the lower branch changes from anti-clockwise to clockwise
direction as the regime changes from unidirectional to a loop, one way
to determine where the transition occurs is to calculate the
parameters at which the mean current $\bar j_{1,4}$ between sites 1
and 4 vanishes.\\
 
As a first step to characterize this change of regime in more detail
we now consider the special case of $\gamma=\delta=0$, which
corresponds to having only injection of particles at the left end of
the lattice and only depletion at the right end. This allows us to
understand the change of regime with fewer parameters and get some
intuition for the cases with non-zero hopping rates $\gamma$ and
$\delta$.\\
 
\begin{figure}[h]
	\begin{center}
		\includegraphics[scale=0.5]{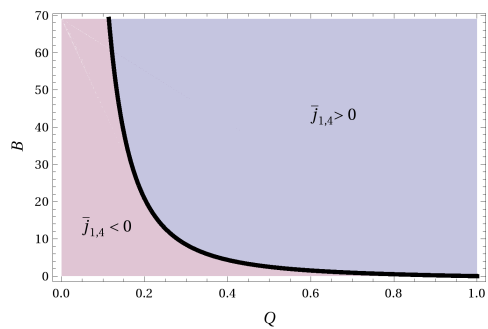}
	\end{center}
	\caption{Regimes of the average current $\bar j_{1,4}$. Red region
$\bar j_{1,4}<0$. Blue region $\bar j_{1,4}>0$}
	\label{ZerCur}
\end{figure}

From equations (\ref{fugacities}) and (\ref{mean_curr}) the condition
for a vanishing mean current $\bar j_{1,4}$ is satisfied when
\begin{equation}
B=\frac{-Q^3+Q^2-Q+1}{Q^2},\label{BofQ}
\end{equation}
where we have defined the ratios $Q=q/p$ and $B=\beta/p$. The curve
defined by the equation above gives us the exact
location of the regime change from positive to negative average
current as shown in figure \ref{ZerCur}. As expected,
if $Q=0$ (i.e., $q=0$) it does not matter how large the extraction
rate $B$ (or $\beta$) is, the system is always
in the loop regime, whereas if $Q=1$ (i.e., for symmetric hopping
rates) the system is always in the unidirectional
regime. Notice also that equation (\ref{BofQ}) has no dependence on
the injection parameter $\alpha$ which means,
that for $0<Q<1$ we can control the system regime just by
changing the extraction rate $\beta$.
For large $\beta$, the system favours the unidirectional regime
whereas, for low $\beta$, a smaller fraction of the
particles on site 3 can leave the system and a loop current is
therefore more likely.
 
In the general case where we admit injection and extraction of
particles from both boundary sites, we can still compute
the mean current $\bar j_{1,4}$:
\begin{equation}
\bar j_{1,4}=\frac{(\alpha+\delta)\left(p^2+q^2\right)\left(q-p\right)+\alpha\beta
q^2-\gamma\delta p^2}
{(p^2+q^2)(\beta+\gamma)+(p+q)\beta\gamma}.\label{fullj14}
\end{equation}
We can check again that for the special cases of symmetric hopping
rates (i.e., $p=q=1$) the current is positive and
the system is in the unidirectional regime, whereas for totally
asymmetric rates (i.e., $p=1$ and $q=0$) the current
is negative and thus, in the loop current regime as expected.
Moreover, for fixed $p>q>0$, we can see that the
mean current regime can be chosen by changing the boundary rates
where, just as in the special case discussed above,
large $\beta$ favours the unidirectional regime and small $\beta$ the
loop current regime. Similarly, for fixed boundary
conditions satisfying $\alpha\beta>\gamma\delta$, we can choose the
mean current regime by varying the bulk hopping
rates $p$ and $q$. A further remark about the existence of these two
regimes, is that if we were to consider $w_n$
bounded we know that for some choices of the hopping rates the system
would undergo condensation (and have no steady
state) meaning, for example, that not all of the phase plane in figure
\ref{ZerCur} would be accessible.
 
\section{Particle current fluctuations}\label{s:currentfluct}
 
\subsection{Large deviations}
 
We now complement our results for the mean currents by studying their
fluctuations. To be specific, we are interested in calculating the
probability distribution $p(j_{a,b},t)$ of the time-averaged current
$j_{a,b}$ between sites $a$ and $b$ in the lattice. In the long-time
limit, we expect this distribution to follow the large deviation
principle
\begin{equation}
 p(j_{a,b},t)\sim \exp\left(-t\hat e(j_{a,b})\right)\label{largedevprinc},
\end{equation}
with rate function (RF) $\hat e(j_{a,b})$. Here ``$\sim$'' denotes
asymptotic equality in the limit of large time. One way to obtain the
RF is to calculate the scaled cumulant generating function (SCGF)
\begin{equation}
e(\lambda)=\lim\limits_{t\rightarrow\infty}-\frac{1}{t}\log\langle
e^{-\lambda tj_{a,b}}\rangle.
\end{equation}
Indeed, if the SCGF is continuous and differentiable, then it is known
that the RF is obtained from the SCGF via the Legendre transform
\cite{Tou09} as
\begin{equation}
\hat e(j_{a,b})=\max\limits_{\lambda}\left\{e(\lambda)-\lambda
j_{a,b}\right\} \label{LegTra}.
\end{equation}
 
To calculate the SCGF we need to modify the Hamiltonian (\ref{HamDia})
to count particle jumps. This is done by multiplying the terms
accounting for the transfer of particles from sites $a$ to $b$ (and
vice versa) by the exponential factor $e^{\mp\lambda}$, see for
example \cite{HarSch07,DerLeb,LebSpo}. Then for instance, to measure the
fluctuations of $\bar j_{1,4}$ the required modified Hamiltonian is:
\begin{eqnarray}
-\hat H=& \alpha\left(a_1^+-1\right) + \gamma\left(a_1^--d_1\right)
  + \delta\left(a_3^+-1\right) + \beta\left(a_3^--d_3\right)\nonumber\\
  &+ \sum\limits_{k=1}^{4}p\left(a_k^-a_{k+1}^+e^{\lambda\delta_{k,4}}-d_k\right)
+ q\left(a_k^+a_{k+1}^-e^{-\lambda\delta_{k,4}}-d_{k+1}\right).
\label{HamDiaMod}
\end{eqnarray}
The SCGF can be identified with the lowest eigenvalue of the
Hamiltonian (\ref{HamDiaMod}), which in general is different from
zero. Then we find the desired RF of the current $j_{1,4}$ via the
Legendre transform (\ref{LegTra}).
 
\subsection{Joint probability of current fluctuations}
 
\begin{figure}[h]
	\begin{center}
		\includegraphics[scale=0.6]{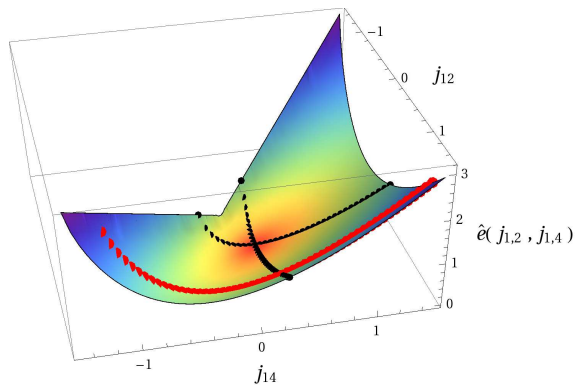}\\
	\end{center}
	\caption{Joint rate function with parameters
$\alpha=1/2$, $G=0$, $\delta=0$, $B=1$ and $Q=1$ (i.e., unidirectional
regime). Black lines cross at the minimum current $j_{1,2}=j_{1,4}=1/4$.
Red line corresponds to the curve with constant $j_{1,2}=5/4$ for which 
the most likely value of $j_{1,4}$ is negative.}
	\label{LegTra2D}
\end{figure}

Correlations between particle currents flowing through different bonds
can be studied by considering a two-parameter SCGF. For the diamond
lattice, it is interesting for example to study the current flowing
between sites 1 and 2 and the current flowing between sites 1 and 4
simultaneously. The joint SCGF that characterizes the joint
distribution $p\left(j_{1,2},j_{1,4},t\right)$ of these currents is
calculated analytically by modifying the system's Hamiltonian
similarly to described above. Particle jumps between sites 1 and 2 are
counted with parameter $\lambda_{1,2}$ and between sites 1 and 4 with
parameter $\lambda_{1,4}$. This calculation leads to the two-parameter
SCGF,
\begin{eqnarray}
	\fl\mathsmaller{e(\lambda_{1,2},\lambda_{1,4})}=
	\mathsmaller{\frac{ \left(Q+1\right)	\left(	\left(\alpha B+\delta
Ge^{\lambda_{1,2}+\lambda_{1,4}}\right) \left(
Q^2e^{\lambda_{1,2}}+e^{\lambda_{1,4}} \right) - \left(\alpha B+\delta
G\right) \left(Q^2+1\right) e^{\lambda_{1,2}+\lambda_{1,4}}	\right) }
	{Q^4 e^{2\lambda_{1,2}}+e^{2\lambda_{1,4}}-e^{\lambda_{1,2}+\lambda_{1,4}}\left(1+Q^4+B\left(Q+1\right)\left(Q^2+1\right)
+G\left(Q+1\right)\left(Q^2+BQ+B+1\right)\right) } }\nonumber\\
	\mathsmaller{+\frac{\left(\alpha+\delta\right)\left(	Q^4e^{2\lambda_{1,2}}-e^{\lambda_{1,2}+\lambda_{1,4}}\left(
Q^4+1 \right)+e^{2\lambda_{1,4}}	\right)}
	{Q^4 e^{2\lambda_{1,2}}+e^{2\lambda_{1,4}}-e^{\lambda_{1,2}+\lambda_{1,4}}\left(1+Q^4+B\left(Q+1\right)\left(Q^2+1\right)
+G\left(Q+1\right)\left(Q^2+BQ+B+1\right)\right) }
}\label{FullJointSCGF}
\end{eqnarray}
where we have defined $Q=q/p$, $B=\beta/p$ and $G=\gamma/p$. For the
special case $\gamma=\delta=0$ the SCGF reduces to
\begin{equation}
	e(\lambda_{1,2},\lambda_{1,4})=\mathsmaller{\alpha\left(1+\frac{B(1+Q)(e^{\lambda_{1,4}}+e^{\lambda_{1,2}}Q^2)}{e^{2\lambda_{1,4}}+e^{2\lambda_{1,2}}Q^4-e^{\lambda_{1,2}+\lambda_{1,4}}(Q^4+1+(1+Q)(1+Q^2)B)}\right)}\label{JointSCGF}.
\end{equation}
The RF $\hat e(j_{1,2},j_{1,4})$ associated with the joint
distribution $p\left(j_{1,2},j_{1,4},t\right)$ is obtained from this
SCGF via the double Legendre transform:
\begin{equation}
	\hat e(j_{1,2},j_{1,4})=\max\limits_{\lambda_{1,2},\lambda_{1,4}}\left\{
e(j_{1,2},j_{1,4})-j_{1,2}\lambda_{1,2}-j_{1,4}\lambda_{1,4}\right\}.
\end{equation}
The numerical evaluation of this transform is shown in figure \ref{LegTra2D}.\\
 
The RF of figure \ref{LegTra2D} is not defined for all currents, since
for $\gamma=\delta=0$ it follows that $j_{1,2}+j_{1,4}\geq0$ in the long time limit. For the
parameters used in figure \ref{LegTra2D}, we also know from our
analysis of section \ref{s:currentloops} that each of the mean
currents $\bar j_{1,2}$ and $\bar j_{1,4}$ corresponding to the minima
of $\hat e(j_{1,2},j_{1,4})$, are strictly positive, which means that
we are in the unidirectional regime. What the joint RF shows is that,
despite being in this regime, it is possible to have loop current
fluctuations. In particular, for a fixed large positive current
$j_{1,2}$, the most probable value of $j_{1,4}$ is negative (as this
minimises the RF), implying that loop current fluctuations are more
likely than unidirectional current fluctuations in this case. This
behaviour is also seen for other values of parameters.\\
 
Note that from the joint SCGF we can quickly recover the SCGF of
either of the two possible partial currents or the total current
simply by making the correct selection of parameters $\lambda_{1,2}$
and $\lambda_{1,4}$. We shall do this in the following subsections.
 
\subsection{Partial current fluctuations}
 
A possible use of the two-parameter SCGF obtained above is to analyse
the partial currents $j_{1,2}$ or $j_{1,4}$. Since the particle flow
from site 1 to site 4 allows us to observe both positive and negative
mean currents (see section 3), we choose to analyse this bond more
closely. To do this we set the parameter $\lambda_{1,2}=0$ and the resulting function is
\begin{equation}
e(\lambda_{1,4})=\mathsmaller{\frac{\left(e^{\lambda_{1,4}}-1\right)\left(
e^{\lambda_{1,4}} \left( G\delta(1+Q)+(\alpha+\delta) \right) -
Q^2\left( \alpha B(1+Q)+Q^2(\alpha+\delta) \right)
\right)}{e^{2\lambda_{1,4}}-e^{\lambda_{1,4}}\left( 1+Q^4+\left(
1+Q+Q^2+Q^3 \right)\left(B+G\right)+BG\left(1+Q\right)^2\right)+Q^4}}.\label{SCGF14}
\end{equation}
 
\begin{figure}[h]
	\begin{center}
		\includegraphics[scale=0.5]{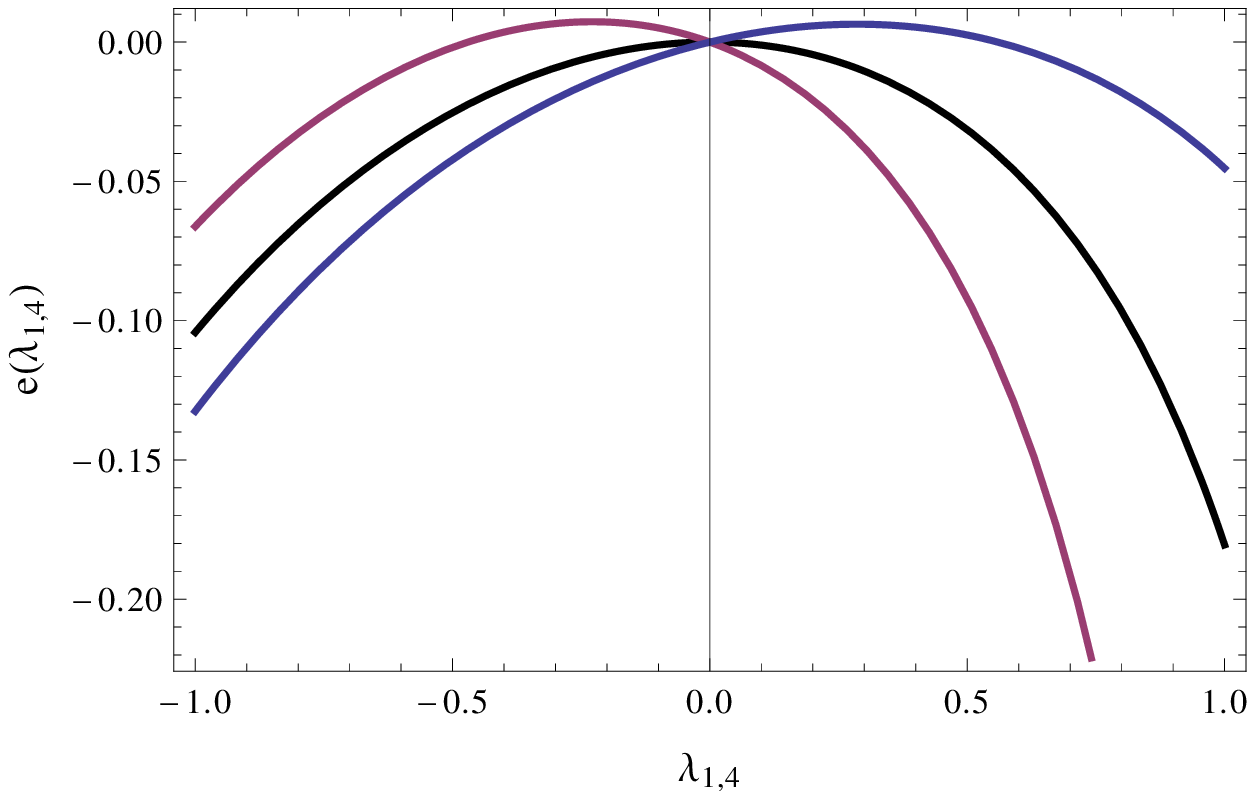}\\
		\includegraphics[scale=0.5]{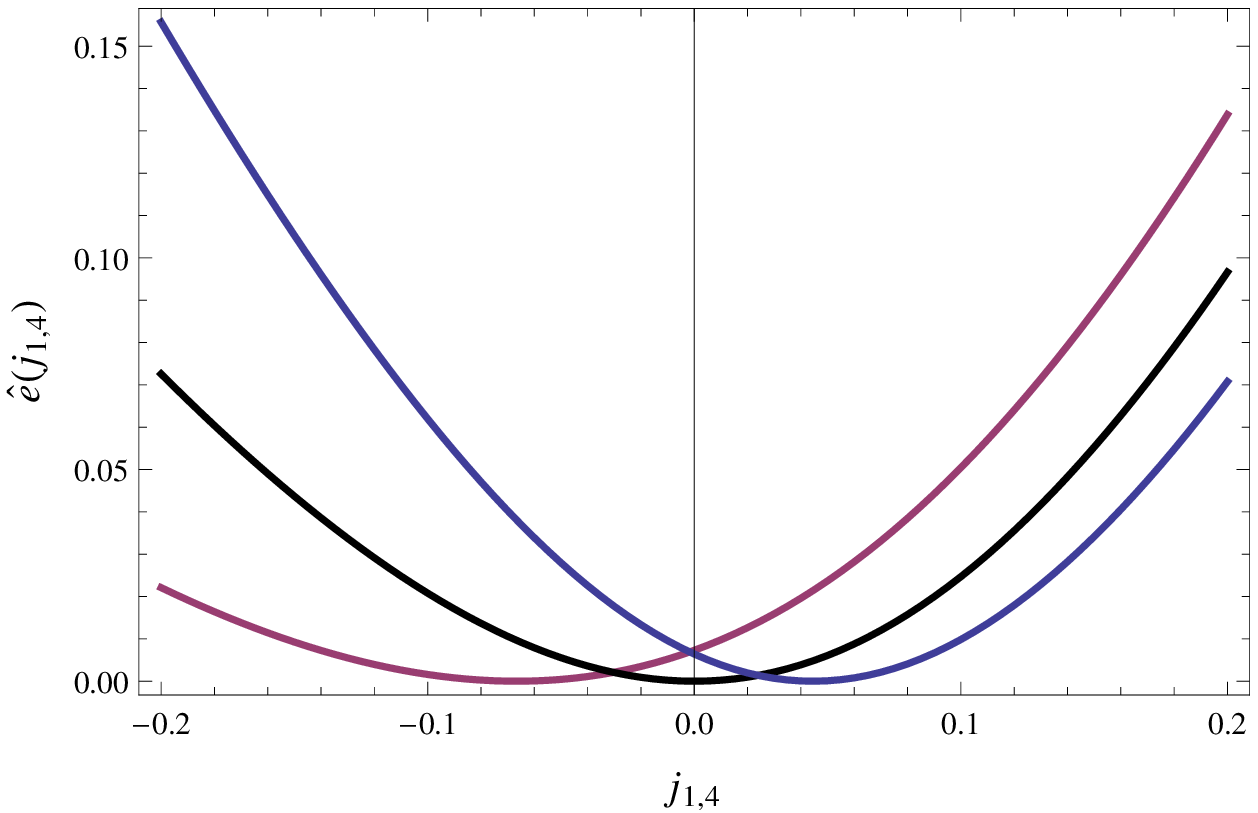}
	\end{center}
	\caption{SCGF (above) and RF (below) for current $j_{1,4}$ with
parameters $\alpha=1/2$, $Q=1/2$ and $B=3/2$ (red), $B=5/2$ (black),
$B=9/2$ (blue). }
	\label{SCGFandRF}
\end{figure}

We can further simplify this function by considering the case
$\gamma=\delta=0$ in which we already studied the change of regime
from unidirectional to loop mean current. The resulting SCGF and the
numerical RF are shown in figure \ref{SCGFandRF}. Here we can observe
again how the choice of the parameter $B$ affects the mean current
flow through the lattice.\\
 
We can see directly from the SCGF whether the choice of parameters
corresponds to a unidirectional or loop regime by looking at the slope
of the function at $\lambda=0$. This is because, from large deviation
theory, we know that the first derivative of the SCGF at zero
corresponds to the mean current between the sites we are looking at.
Hence, a positive slope implies a positive current and the
unidirectional regime, whereas a negative slope implies the loop
regime. At the level of the RF, this translates into having a positive
or negative minimum, respectively, which determines the mean current.
From the full form of the RF, what can be seen again is that
fluctuations of currents having a sign opposite to the mean current
have a non-zero probability to be observed. This means concretely that
loop current fluctuations can be seen in the unidirectional current
regime and vice versa.
 
\subsection{Total current fluctuations}
 
Another possibility, is not to differentiate between the two lattice
branches but simply to measure the total flux of particles through the
cross-section between site 1 and sites 2 and 4, i.e.,
$j=j_{1,2}+j_{1,4}$. Once again, the joint SCGF (\ref{FullJointSCGF})
reduces to a single variable function but now by taking $\lambda_{1,2}
=\lambda_{1,4} =\lambda$:
\begin{eqnarray}
	e(\lambda)=\frac{\left(1-e^{-\lambda}\right)
\left(Q^2+1\right)\left(	\alpha B - \delta
Ge^{\lambda}	\right)}{B\left(Q^2+1\right)+G\left(Q^2+BQ+B+1\right)}.\label{SCGF_samelambda}
\end{eqnarray}
The RF $\hat e(j)$ associated with the probability distribution of $j$
can be obtained analytically for this SCGF, resulting in
\begin{eqnarray}
\hat e(j)=&\mathsmaller{-\frac{-\left(p^2+q^2\right) (\alpha  \beta +\gamma  \delta )}{(p^2+q^2) (\beta +\gamma )+(p+q) \beta  \gamma}\nonumber}\\
&\mathsmaller{+\frac{\sqrt{\left(j \left(p^2+q^2\right) \beta +j \left(p^2+q^2+(p+q) \beta \right) \gamma \right)^2+4 \left(p^2+q^2\right)^2 \alpha  \beta  \gamma  \delta }}{(p^2+q^2) (\beta +\gamma )+(p+q) \beta  \gamma}\nonumber}\\
&\mathsmaller{+\frac{j \left(p \beta  \gamma +p^2 (\beta +\gamma )+q (q \beta +(q+\beta ) \gamma )\right)}{(p^2+q^2) (\beta +\gamma )+(p+q) \beta  \gamma}\nonumber}\\
&\mathsmaller{\times \log\left( \frac{-j \left(p \beta  \gamma +p^2 (\beta +\gamma )+q (q \beta +(q+\beta ) \gamma )\right)}{2 \left(p^2+q^2\right) \gamma  \delta} \right. \nonumber}\\
&\mathsmaller{\left.\quad+\frac{\sqrt{\left(j \left(p^2+q^2\right) \beta +j \left(p^2+q^2+(p+q) \beta \right) \gamma \right)^2+4 \left(p^2+q^2\right)^2 \alpha  \beta  \gamma  \delta }}{2 \left(p^2+q^2\right) \gamma  \delta }\right)}. \label{DiamRF}
\end{eqnarray}
As expected, for the case of unbounded $w_n$ the same RF is found when
measuring current fluctuations of $j_{2,3}+j_{4,3}$ or either of the
boundary currents $j_L$ or $j_R$. We shall use equation (\ref{DiamRF})
in the next subsection to show that fluctuations of total currents
possess a particular symmetry property that fluctuations of partial
currents do not have.
 
\subsection{Fluctuation symmetries}
 
In the previous subsections we have seen that the large deviation RF
provides rich information about stochastic models beyond the level of
mean values. The study of such fluctuations allows us to find
symmetries in the dynamics of a system rather than just spatial
symmetries. In particular, the Gallavotti-Cohen fluctuation relation
(GCFR) provides a well-known symmetry for the entropy production rate,
which can be translated into a symmetry for the current RFs in
interacting particle systems
\cite{HarSch07,LebSpo,EvaCohMor,GalCoh}.\\
 
The GCFR in the context of particle current fluctuations can be written as
\begin{equation}
\frac{p(-j,t)}{p(j,t)}\sim e^{-Ejt}\label{GCFTrel}
\end{equation}
and makes a connection between the long-time probability of observing
a time-averaged current $j$ and the probability of observing a current
$-j$ of the same magnitude in the reversed direction. Here, $E$ is a
constant which can be interpreted as an equilibrium-restoring
field. Since we are assuming that currents follow a large deviation
principle as in equation (\ref{largedevprinc}), relation
(\ref{GCFTrel}) can also be written in terms of only the RFs as
\begin{equation}
\hat e(-j)-\hat e(j)=Ej.\label{linearGCFT}
\end{equation}
 
Now that we know the probability rate function of observing particle
currents measured through different bonds, we would like to
investigate if our rate functions satisfy the GCFR. In our model it
can be straightforwardly shown analytically that the RF (\ref{DiamRF})
for the fluctuations of the \emph{total} current, $j=j_{1,2}+j_{1,4}$,
obeys the GCFR (\ref{linearGCFT}) with
\begin{equation}
	E=\log\left(\frac{\alpha\beta}{\gamma\delta}\right).
\end{equation}
Interestingly, due to the cyclic arrangement of the bulk hopping rates
in this model, the equilibrium-restoring field $E$ turns out to be
independent of the rates $p$ and $q$ and equal to the expression for a
single-site ZRP with open boundaries \cite{HarRakSch06}.\\
 
\begin{figure}[h]
	\begin{center}
		\includegraphics[scale=0.5]{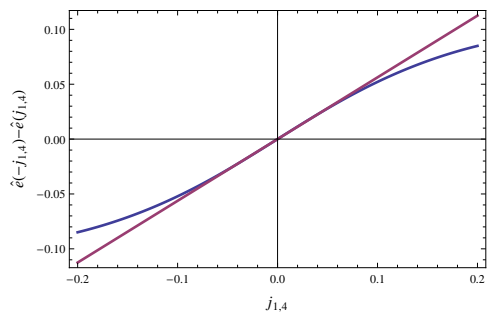}\\
	\end{center}
	\caption{GCFR breakdown for partial current $j_{1,4}$. The blue curve
corresponds to the numerical values of $\hat e(-j_{1,4})-\hat
e(j_{1,4})$ with parameters $Q=1/2$, $\alpha=1/2$, $G=\delta=0$ and
$B=9/2$ (unidirectional regime) whereas the straight line corresponds
to $\hat e(-j_{1,4})-\hat e(j_{1,4})=E j_{1,4}$ with field $E=0.563$.}
	\label{GCFT}
\end{figure}

On the other hand, for \emph{partial} current fluctuations, the GCFR
(\ref{linearGCFT}) is \emph{not} satisfied. This can be clearly seen
in figure \ref{GCFT} where we plot $\hat e(-j_{1,4})-\hat e(j_{1,4})$
for the numerical results obtained in section 4.3 and observe that is
not linear in $j_{1,4}$. However, we can see that for small $j_{1,4}$
the relation is approximately satisfied.\\
 
\begin{figure}[h]
	\begin{center}
		\includegraphics[scale=0.7]{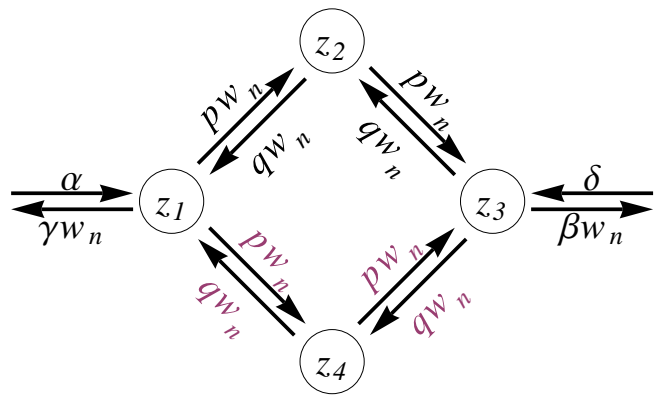}
	\end{center}
	\caption{Model variant with reversed hopping rates on lower branch.}
	\label{SqrLat}
\end{figure}

Our observations are consistent with recent results for other two
dimensional models, both classical and quantum, which highlight the
breakdown of the GCFR for partial currents \cite{BodDerLeb,KraSchBra}\footnotemark.
We emphasize here that the crucial point is not the existence of a
loop mean current regime but that for some partial current
fluctuations away from the mean, the most likely realisation involves
loop current flow. This is confirmed by studying a variant of our
model in which the rates on the lower branch are reversed (see figure
\ref{SqrLat}) and thus the possibility of a loop mean current is
eliminated. Performing a similar analysis as for the original diamond
model one again observes loop fluctuations in the joint-probability RF
and finds that the GCFR is satisfied for the total current, this time
with $E=\log\frac{p^2\alpha\beta}{q^2\delta\gamma}$, but
not for partial currents.\\

\footnotetext{Note that all these cases involve the splitting of the current into
different branches; they are different to the initial-condition-related
GCFR breakdown for certain currents observed by Visco \cite{Vis} and others.}
 
We conclude this subsection by commenting on the relation of our results 
to recent general analysis in the literature. A theory for electron 
currents on arbitrary networks was presented in \cite{GanSin} and confirmed 
for some specific cases; our model provides a complementary classical 
example in which fluctuations can be calculated exactly for interacting 
particles with \emph{any} unbounded interaction $w_n$.  Note that the total 
current $j$ in our system flows through input and output links which in the 
language of \cite{GanSin} ``do not belong to any loop of the 
network'' and therefore the observed GCFR for the total current is consistent 
with the general arguments there. However, for the original version of 
the diamond model in figure \ref{DiaArr}, this total current $j$ is \emph{not} proportional to the 
entropy production\footnotemark. The possibility of 
a GCFR-type symmetry for quantities other than entropy was also highlighted 
in \cite{BarCheHinMuk} in a slightly different context -- a ``peculiar" 
network of configurations in state space.  Our model provides a simple 
demonstration of this newly proposed symmetry but for interacting particles 
on a real-space network and corresponding infinite state space. Note, however, 
that for the modified model of figure \ref{SqrLat} the total current $j$ is proportional to the entropy production rate.
\footnotetext{It can readily be shown that the time-averaged 
entropy production in this case consists of a contribution from current 
around the loop as well as a term proportional to $j$.}

\section{Diamond chain}\label{s:diamondchain}
 
\begin{figure}[h]
	\begin{center}
		\includegraphics[scale=1]{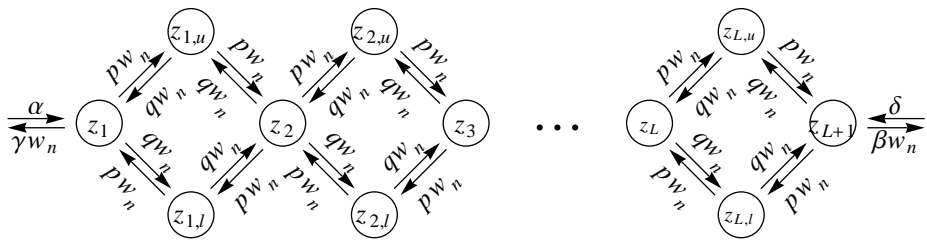}
	\end{center}
	\caption{Chain of $L$ diamonds. Here the $m$th-diamond is formed by 
		sites $z_m$, $z_{m+1}$, upper branch site $z_{m,u}$ and lower branch
		site $z_{m,l}$.}
	\label{DiamondChain}
\end{figure}

We illustrate the generality of our approach by extending the analysis to a 
chain of coupled diamonds as shown in figure \ref{DiamondChain}. 
Following the discussion in section
\ref{s:currentloops}, for partially
asymmetric hopping rates $p\neq q$ we expect the mean current regime in the 
different diamonds to depend on the boundary parameters. Furthermore, we now 
have the possibility of co-existence of unidirectional and loop current regimes 
along the chain -- a scenario we shall explore in more detail below.\\

From calculation of the mean steady-state currents as in section 2.2, we find that to see a rightwards 
total mean current through the system, the boundary parameters must again satisfy 
\begin{equation}
	\alpha\beta>\gamma\delta \label{cond3}.
\end{equation}
Assuming this condition, we now concentrate on determining how tuning the extraction rate, 
$\beta$, controls the presence or absence of mean current loops for each diamond. 
From our analysis of the single diamond array, we 
expect that for a given asymmetry, small $\beta$ favours the 
loop current regime, whereas large $\beta$ leads to the unidirectional regime. Furthermore, 
one intuitively expects that for large chains the influence of $\beta$ in the diamonds close 
to the left-hand side boundary is small. Therefore, one expects for the diamond chain that 
for increasing $\beta$, the diamonds close to the right-hand side boundary change to the 
unidirectional regime before the ones close to the opposite boundary. 
In other words, for intermediate $\beta$ there should be co-existence of two current 
domains located at left and right sides of the chain.\\

A detailed analysis of the mean currents in the lower branch of each diamond confirms the above picture. 
However, it turns out that for strongly asymmetric bulk rates, the left-hand diamonds remain in the loop current 
regime regardless of how large the extraction rate is. For $p>q$, considering the behaviour of the first diamond 
in the chain we can show that in order to find a finite $\beta$ 
such that all diamonds change regime, the hopping rates must satisfy
\begin{equation}
	\left(\frac{q}{p}\right)^2>1-\frac{1}{L},
\end{equation}
where $L$ is the number of diamonds in the chain. Hence for weakly asymmetric hopping rates, there is a 
crossover between a phase in which there is a loop current in every diamond and a phase where the current 
is everywhere unidirectional. This transition persists even in the thermodynamic limit, 
$L\rightarrow\infty$.\\

\begin{figure}[h]
	\begin{center}
		\includegraphics[scale=0.8]{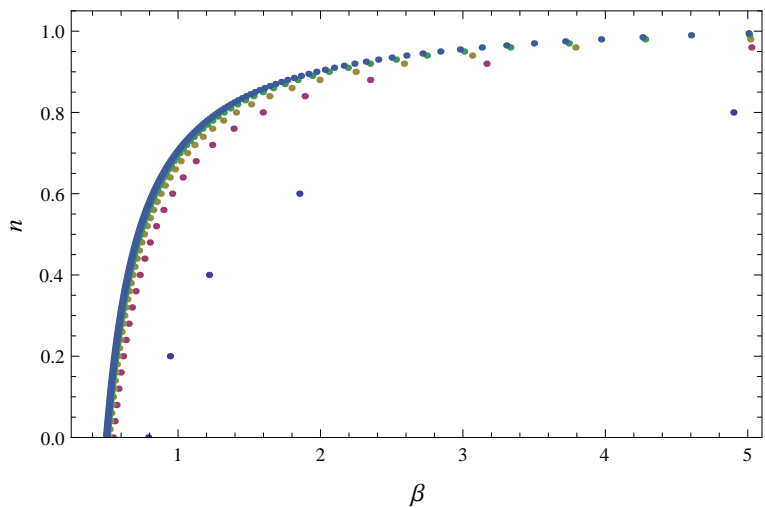}
	\end{center}
	\caption{Proportion of diamonds in unidirectional 
	regime as function of $\beta$ for $p=1$, $\alpha=1$, $\gamma=1$, $\delta=1/4$, $\xi=19/20$ and $L=$ 5, 
	25, 50, 100, 200 from bottom to top.}
	\label{UnidDiamRatio}
\end{figure}

Focusing on this weakly asymmetric case, we now set
\begin{equation}
	\frac{q}{p}=1-\frac{\xi}{2L},
\end{equation}
where $\xi$ takes values in $(0,1)$. Using exact numerics to check the mean current in each diamond, 
we plot in figure \ref{UnidDiamRatio} how the proportion $n$ of diamonds in the unidirectional current 
regime increases with the parameter $\beta$. 
One can clearly see convergence towards a limiting curve as $L$ increases. In the thermodynamic 
limit, the exact critical value at which the right most diamond changes regime is given analytically by
\begin{equation}
	\beta_c=\frac{\gamma\delta}{\alpha}\left(1+\xi\right).
\end{equation}
Note that this critical point always satisfies assumption (\ref{cond3}) which is essential for a 
rightwards total mean current.\\

We can also calculate the generating function for the current fluctuations through any 
of the cross-sections of the chain by using the same method as in section \ref{s:currentfluct}. 
In this case, we verify again that it is only for total currents that the GCFR is satisfied; specifically, 
we obtain the same field $E$ as for the single diamond regardless 
of the length of the chain.

\section{Conclusion}\label{s:conclusion}
 
We have performed a current fluctuation analysis of the zero-range
process on a diamond lattice with open boundaries.
For rates which impose a preferred direction around the diamond, we
demonstrated the possibility of two
different mean current regimes depending on the boundary parameters
and the asymmetry of the bulk hopping
rates. Moreover, for the case in which particles are only injected on
one side of the diamond and only removed from
the other we proved that, regardless of the injection rate, we could
control the regime of the system just by varying
the extraction rate.\\
 
Analysing the current fluctuations via the two-parameter scaled
cumulant generating function, we also studied the joint
probability distribution of the partial and total currents flowing
between sites 1 and 2 and between sites 1 and 4 of
the lattice. Significantly, we saw from the joint current rate
function that both unidirectional and loop current
fluctuations may be observed whether the mean current is
unidirectional or in a loop.\\
 
From our analysis  we also confirmed that to observe the
Gallavotti-Cohen fluctuation relation we need to measure total
currents as opposed to partial currents. This point should obviously be taken into account when
testing fluctuation symmetries in higher dimensional
systems in experiment or simulation. Indeed, the recent isometric
fluctuation relation \cite{HurPer} which is a generalised
symmetry for higher dimensions, is also concerned with global rather
than local currents.\\
 
Finally, we 
applied our methods to a chain of diamonds and explicitly demonstrated 
that it can support co-existence of different current domains. We 
emphasize that our results hold only for the case of an unbounded
site interaction $w_n$; it would be interesting to extend the analysis
to the case of bounded $w_n$ where one expects the formation of
``instantaneous condensates'' and a breakdown of the fluctuation
symmetry even for total currents
\cite{HarRakSch06,RakHar08,HarRakSch05}. For such models on higher
dimensional lattices or more complex geometries we
expect loop current or vortex fluctuations to play an important role.
 
\section*{References}

\end{document}